# Localization of Simultaneous Multiple Sources using SMS-LORETA

Avni Pllana[*] and Herbert Bauer

Biological Psychology Unit

Dept. of Clinical, Differential and Biological Psychology

Vienna University

Liebiggasse 5, 1010 Vienna, Austria.

---

[*] Corresponding author, E-mail: avni.pllana@univie.ac.at, Fax: ++4314277-47939

# Localization of Simultaneous Multiple Sources using SMS-LORETA

Avni Pllana and Herbert Bauer

**Abstract**:

In this paper we present a new localization method SMS-LORETA (Simultaneous Multiple Sources- Low Resolution Brain Electromagnetic Tomography), capable to locate efficiently multiple simultaneous sources. The new method overcomes some of the drawbacks of sLORETA (standardized Low Resolution Brain Electromagnetic Tomography). The key idea of the new method is the iterative search for current dipoles, harnessing the low error single source localization performance of sLORETA. An evaluation of the new method by simulation has been enclosed.

**Key words**: EEG, MEG, Inverse problem, SMS-LORETA, sLORETA, Iteration.

**Introduction**

A large number of simulations performed by different authors (Wagner et al., 2004; Soufflet and Boeijinga, 2005), confirmed the low error localization performance of sLORETA introduced by Pascual-Marqui in 2002 (Pascual-Marqui, 2002), in the case of single current dipoles as sources. However, it has been noted that a low localization error for single dipoles is not a sufficient condition for the correct localization of multiple simultaneous sources (Wagner et al., 2004; Soufflet and Boeijinga, 2005). Our motivation was to overcome this drawback and develop a more powerful localization method, and the result of our efforts is the new method SMS-LORETA, presented below.

*Preliminaries*

In this section we introduce some basic definitions concerning sLORETA. The forward EEG (electroencephalography) equation is as follows

$$\mathbf{\Phi} = \mathbf{KJ} + c\mathbf{1} \qquad (1)$$

where the vector $\mathbf{\Phi} \in R^N$ represents the scalp electric potential differences measured by N electrodes with respect to a reference electrode. $\mathbf{K} \in R^{N \times 3M}$ is the so-called lead field matrix representing the gain of each current dipole placed in a brain voxel, to every extra-cranial electrode; M is the number of predefined points (voxels) in the brain volume, equally spaced and forming a cubic grid. This set of M points is called the solution space. Matrix $\mathbf{K}$ has 3M columns, because every current dipole is a 3D vector. Constant c is a scalar reflecting the physical fact that electric potential can be determined up to an arbitrary constant; and $\mathbf{1} \in R^N$ is a vector of ones. $\mathbf{J} \in R^{3M}$ is the vector of current dipole moments.

In order to obtain results that are independent of the reference electrode, the EEG measurements and the lead field are transformed as follows

$$\mathbf{\Phi} \leftarrow \mathbf{H}\mathbf{\Phi},$$

$$\mathbf{K} \leftarrow \mathbf{H}\mathbf{K},$$

where $\mathbf{H} = \mathbf{I} - \dfrac{\mathbf{1} \cdot \mathbf{1}^T}{\mathbf{1}^T \cdot \mathbf{1}}$, is called the average reference operator, or the centering matrix, and $\mathbf{H} \in R^{N \times N}$ (see for details: Pascual-Marqui, 1999, 2002, 2007).

Now equation (1) has the form

$$\mathbf{\Phi} = \mathbf{K}\mathbf{J}. \qquad (2)$$

The inverse problem can be stated as follows: From given scalp potential measurements $\mathbf{\Phi}$ determine the corresponding current density distribution $\hat{\mathbf{J}}$ generated by the neuronal activity. This problem has no unique solution, but we can find a minimum norm estimate

$$\hat{\mathbf{J}} = \mathbf{T}\mathbf{\Phi}, \qquad (3)$$

where $\mathbf{T} = \mathbf{K}^T\left[\mathbf{K}\mathbf{K}^T + \alpha\mathbf{H}\right]^+$, $\alpha \geq 0$ is a regularization parameter needed to overcome numerical problems with the ill-conditioned matrix $\mathbf{K}$. The superscript $^+$ denotes the Moore-Penrose pseudoinverse. According to sLORETA the estimate $\hat{\mathbf{J}}$ should be standardized using the variance of the estimated current density

$$\mathbf{S}_{\hat{\mathbf{j}}} = \mathbf{K}^T\left[\mathbf{K}\mathbf{K}^T + \alpha\mathbf{H}\right]^+\mathbf{K}. \qquad (4)$$

The estimated current density is then

$$\overline{\mathbf{J}}_l = \left\{\left[\mathbf{S}_{\hat{\mathbf{j}}}\right]_{ll}\right\}^{-1/2} \hat{\mathbf{J}}_l, \qquad (5)$$

where $\hat{\mathbf{J}}_l \in R^3$ is the estimated current dipole at location $l$ according to equation (3), and $\left[\mathbf{S}_{\hat{\mathbf{j}}}\right]_{ll} \in R^{3 \times 3}$ is the $l^{th}$ diagonal block of matrix $\mathbf{S}_{\hat{\mathbf{j}}}$. In order to simplify the notation we define the following matrix

$$\mathbf{S}_d = \begin{bmatrix} \left[\mathbf{S}_{\hat{\mathbf{j}}}\right]_{11}^{-1/2} & 0 & \cdots & 0 \\ 0 & \left[\mathbf{S}_{\hat{\mathbf{j}}}\right]_{22}^{-1/2} & \cdots & 0 \\ \vdots & \vdots & \ddots & 0 \\ 0 & 0 & 0 & \left[\mathbf{S}_{\hat{\mathbf{j}}}\right]_{MM}^{-1/2} \end{bmatrix}. \qquad (6)$$

**Methods**

In the case of simultaneous multiple sources we can use the outcome of sLORETA as initial guess for an iterative search of active sources. Let $\hat{\mathbf{J}}_{i_0}$ be the maximum strength dipole computed by sLORETA, where the voxel index $i_0 \in \{1, 2, \ldots, M\}$. Using equation (2) we determine the contribution $\mathbf{F}_b$ of $\hat{\mathbf{J}}_{i_0}$ to the measured potential distribution $\mathbf{\Phi}$. We subtract $\mathbf{F}_b$ from $\mathbf{\Phi}$ and employ again sLORETA to determine a new maximum strength dipole $\hat{\mathbf{J}}_{i_1}$ for the depleted potential distribution $\mathbf{\Phi} - \mathbf{F}_b$. We continue this negative feedback as long as the norm of the depleted potential is larger than 5% of the norm of the measured potential distribution. We store the voxel index of the estimated maximum strength dipole of each iteration. After termination of the negative feedback, we have a vector of voxel indices of the locations of the estimated dipoles that occurred. This vector is then sorted in descending order according to the occurrence multiplicity of the voxel indices. The multiplicity of occurrence is proportional to the strength of the true current dipoles. Therefore the first few sorted locations are the probable locations of the strongest true current dipoles. The SMS-LORETA algorithm is illustrated by the following pseudocode:

$\mathbf{F} = \mathbf{\Phi}$  /∗ Measured potential distribution

$k = 0$

$\left|\hat{\mathbf{J}}_{i=i_0}\right| = \max_{i=1,\ldots,M}\left|\hat{\mathbf{J}}_i\right|$  /∗ Initial guess computed by sLORETA

$m = i_0$

while $|\mathbf{F}| > 0.05\,|\mathbf{\Phi}|$

    $k = k + 1$

    $\mathbf{F}_b = \mathbf{K}_m \hat{\mathbf{J}}_m$  /∗ The use of Eq. (2)

    $\mathbf{F} = \mathbf{F} - \mathbf{F}_b$  /∗ The negative feedback

    $\hat{\mathbf{J}} = \mathbf{T}\mathbf{F}$  /∗ The first step of sLORETA, Eq. (3)

    $\hat{\mathbf{J}} = \mathbf{S}_d \hat{\mathbf{J}}$  /∗ The second step of sLORETA, Eq. (5)

    $\left|\hat{\mathbf{J}}_{i=i_k}\right| = \max_{i=1,\ldots,M}\left|\hat{\mathbf{J}}_i\right|$  /∗ Maximum strength dipole

    $m = i_k$

    $\mathbf{I}(k) = i_k$  /∗ Indices of occurred locations (voxels)

end

Vector $\mathbf{I}$ is the vector of indices of occurred locations. We compute the vector $\mathbf{I}_s$ of voxel indices in $\mathbf{I}$ sorted in descending order according to the multiplicity of their occurrence in $\mathbf{I}$. By means of $\mathbf{I}_s$ we obtain the matrix $\mathbf{L}_n = \mathbf{L}(\mathbf{I}_s(1,\cdots,n))$ of locations of the $n$ strongest estimated current dipoles, where $\mathbf{L} \in R^{M \times 3}$ is the matrix of Cartesian coordinates of M grid points of the solution space.

For simulation purposes the lead field matrix $\mathbf{K}$ can be computed using the spherical 1-shell head model (Fender, 1987), according to which the potential distribution on the sphere is

$$V = \frac{\mathbf{j}}{4\pi\sigma} \cdot \left[ \frac{\hat{\mathbf{x}} - \hat{\mathbf{d}}}{xd(1 - \hat{\mathbf{x}} \cdot \hat{\mathbf{d}})} - \frac{2\hat{\mathbf{d}}}{d^2} \right], \qquad (7)$$

where

- bold-faced type indicates vector quantities, italic type indicates the magnitude of the vector, and hatted variables are unit vectors,
- $\mathbf{j} = [j_x \ j_y \ j_z]$ is the current dipole moment,
- $\sigma$ is the brain conductivity,
- $\mathbf{x}$ is vector from the center of the sphere to the sensor on the surface, that means the scalp electrode,
- $\mathbf{d}$ is the vector from the sensor to the current dipole.

A more realistic head model for computing the $\mathbf{K}$ matrix is the 3 sphere shell model (Phillips et al., 2007). It comprises three concentric spheres of brain radius $r_1$, skull radius $r_2$, and scalp radius $R$. A current dipole $\mathbf{j}$ located at height z on the $\mathbf{e}_z$ axis, generates at the scalp sensor $\mathbf{s} = \mathbf{s}(\theta,\varphi)$ the potential

$$V(\mathbf{s}(\theta,\varphi)) = \frac{1}{4\pi\sigma R^2} \sum_{n=1}^{\infty} \frac{2n+1}{n} f^{n-1} \left[ \frac{\xi(2n+1)^2}{d_n(n+1)} \right] \\ \cdot \left[ (j_x \cos\varphi + j_y \sin\varphi) P_n^1(\cos\theta) + j_z n P_n(\cos\theta) \right], \qquad (8)$$

where

- $f = z/R$ is the eccentricity of the dipole,
- $j_x$, $j_y$, and $j_z$ are components of the dipole $\mathbf{j} = [j_x \ j_y \ j_z]$,
- $\xi = \sigma_{sk}/\sigma$ is the relative conductivity of the skull to the conductivity of the brain and scalp,

- $P_n(\cos\theta)$ and $P_n^1(\cos\theta)$ are Legendre and associated Legendre polynomials. The MATLAB implementation of $P_n^1(\cos\theta)$ is equal to $(-1)P_n^1(\cos\theta)$,

- $d_n = [(n+1)\xi + n]\left[\dfrac{n\xi}{n+1} + 1\right] + (1-\xi)[(n+1)\xi + n](f_1^{2n+1} - f_2^{2n+1}) - n(1-\xi)^2\left(\dfrac{f_1}{f_2}\right)^{2n+1}$,

  where $f_1 = r_1/R$ and $f_2 = r_2/R$.

**Results**

The performance of SMS-LORETA has been evaluated with 1000 noise free simulations, where in each simulation 2 dipoles with random locations and random moments have been located in the solution space. The solution space was a regular grid of points (spacing $dv = 1 cm$) within a volume that was confined by two hemispheres with radius $R_1 = 5 cm$ and $R_2 = 8 cm$, respectively. The number of grid points was M = 812. The number of sensors was N = 91, and the sensors were placed at the nodes of a triangular tessellation of the hemisphere with the radius $R = 9.2\ cm$, as shown in Fig. 1.

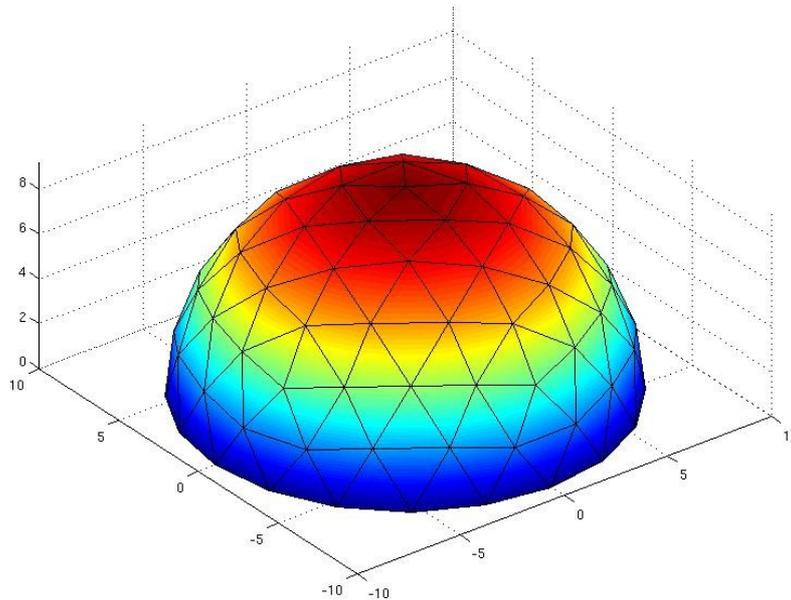

Figure 1: 91 sensor locations at the nodes of triangular tessellation of the simulation hemisphere with radius $R = 9.2\ cm$.

The SMS-LORETA method was implemented in MATLAB, and the **K** matrix was computed using formula (8).

Following numerical values were used:

$r_1$ = 8.0 $cm$ for the brain radius,

$r_2$ = 8.5 $cm$ for the skull radius,

$R$ = 9.2 $cm$ for the scalp radius,

$\sigma$ = 0.33 $(\Omega\, m)^{-1}$ for the brain conductivity, which was assumed to be equal to the scalp conductivity,

$\sigma_{sk}$ = 0.0042 $(\Omega\, m)^{-1}$ for the skull conductivity.

The number of terms of the infinite series in (8) has been limited to n = 50 and the regularization parameter has been chosen $\alpha$ = 0. The simulation results are summarized in Table 1.

|  | sLORETA zero error localization in [%] | SMS-LORETA zero error localization in [%] |
| --- | --- | --- |
| 2 random sources | 1.6 | 56.2 |
| at least 1 of 2 random sources | 81.8 | 79.8 |

Table 1: Localization results for a series of 1000 noise free simulations with 2 simultaneously active dipoles with random locations and random moments.

**Discussion**

Table 1 shows that SMS-LORETA was significantly better than sLORETA in localizing 2 simultaneously active sources. Noise free simulations and generating dipoles placed at the nodes of the cubic grid in the solution space were used, in order to obtain zero error localization, which appeared to be more appropriate for evaluation and comparison. A similar head model with 148 electrodes and 818 grid points was used by Pascual-Marqui (1999). In the simulations presented two simultaneous sources were used, but the proposed algorithm performs well also for a larger number of sources. Formula (8) has been included in the paper in order to support and encourage colleagues to reproduce the results presented. Also a small correction of (8) was made by adding $R^2$ in the denominator of the pre-multiplying factor missing in recent literature (e.g., Phillips et al., 2007). With this correction formula (8) yields results that are in agreement with formula (7) in the case of a 1-shell head model. The extension of results for MEG (magnetoencephalography) is straightforward. Further work will be done to improve the performance of SMS-LORETA, i.e. implementing a realistic head geometry and the Boundary Element Method (BEM) .

**Acknowledgements**

This research was supported by the Austrian Research fund (FWF) P # 19830-B02 and the Austrian National Bank (OeNB) P # 12475.